\documentclass[12pt]{article}
\usepackage{amsmath}
\usepackage{graphicx,psfrag,epsf}
\usepackage{enumerate}
\usepackage[round]{natbib}
\usepackage{url} 
\usepackage{hyperref}
\usepackage{amsmath}
\usepackage{amsthm}
\usepackage{amssymb}
\usepackage{mathrsfs}
\newtheorem{theorem}{Theorem}[section]

\newtheorem{assumption}{Assumption}

\newtheorem{remark}{Remark}[section]

\usepackage{array}
\usepackage{booktabs} 
\usepackage{caption}
\usepackage{tabularx}
\usepackage{float}
\usepackage{graphicx}
\usepackage{subcaption}
\usepackage{caption}
\usepackage{appendix}
\usepackage{multirow}
\usepackage[linesnumbered,ruled]{algorithm2e}

\newcommand{\blind}{1}

\hypersetup{
  colorlinks,       
  citecolor=blue,   
  linkcolor=blue,   
  pdfborder={0 0 0} 
}

\begin{document}

\def\spacingset#1{\renewcommand{\baselinestretch}%
{#1}\small\normalsize} \spacingset{1}


\if1\blind
{
  \title{\bf An Improved Minimization Method: Eliminating Shift and Variance Inflation While Maintaining Marginal Balance}
  \author{\small{Yuhang Tao\textsuperscript{1} and Li-Xin Zhang\textsuperscript{2,3,}}\thanks{Corresponding author. Email address: stazlx@mail.zjgsu.edu.cn} \hspace{.2cm}\\
    \small{\textsuperscript{1}School of Mathematical Sciences, Zhejiang University, Hangzhou, China} \\
    \small{\textsuperscript{2}School of Statistics and Mathematics, Zhejiang Gongshang University, Hangzhou, China} \\
    \small{\textsuperscript{3}Center for Data Science, Zhejiang University, Hangzhou, China} }
  \date{}
  \maketitle
} \fi

\if0\blind
{
  \bigskip
  \bigskip
  \bigskip
  \begin{center}
    {\LARGE\bf }
\end{center}
  \medskip
} \fi

\bigskip
\begin{abstract}
\linespread{1.5}\selectfont
Pocock and Simon’s minimization method is a widely used approach for covariate-adaptive randomization in clinical trials. However, it can exhibit the shift problem under unequal allocation and variance inflation of the imbalances in additional covariates. In this paper, we propose a new minimization method that preserves the conventional biased-coin allocation probability while ensuring marginal imbalances bounded in probability. Meanwhile, the proposed procedure guarantees that neither the shift problem nor variance inflation occurs. Simulation and empirical studies demonstrate the excellent performance of the proposed procedure. 
\end{abstract}

\noindent%
\vfill

\newpage
\spacingset{1.45} 
\section{Introduction}

Covariate-adaptive randomization (CAR) procedures are widely used in clinical trials to balance covariates. Common CAR procedures begin by stratifying patients based on important prognostic covariates and treatments are then assigned to maintain treatment balance within each stratum. Representative examples include stratified permuted block randomization (\citealp{ZELEN1974365}) and stratified urn randomization (\citealp{wei1978}). When the number of covariate strata is large, the above procedures may become less effective. As a result, Pocock and Simon's minimization procedure (PS) (\citealp{taves1974minimization}, \citealp{pocock1975sequential}) provides an alternative, which aims to balance treatment assignments across covariate margins. As reviewed by \cite{TAVES2010180} and \cite{coart2023minimization}, the PS procedure has been increasingly used in clinical trials for its ability to improve covariate balance. In recent years, significant progress has been made in understanding its theoretical properties, which has further promoted its application (\citealp{hu2020theory}, \citealp{hu2023multi}). 

Despite its widespread adoption, the PS procedure has two shortcomings, which have led to some concerns. First, using simulation, \cite{zhao2024consistent} showed that under the PS procedure, the limiting covariance matrix of within-stratum imbalances is not necessarily dominated by that under complete randomization (CR), and in some directions, the variance is larger under the PS procedure than under CR. Consequently, the conventional Wald test for the average treatment effect can suffer from asymptotic Type I error inflation under the PS procedure. Second, when PS is extended to unequal allocation ratios (\citealp{liu2017general}), simulation studies by \cite{fang2026general} suggested that it may suffer from the shift problem, that is, the within-stratum imbalance, divided by the sample size, converges in probability to a nonzero constant. As a consequence, the difference-in-means estimator for the average treatment effect may be biased, thereby reducing the validity and power of the resulting inference. As early as 2015, \cite{EMA2015guideline} had already raised relevant concerns and pointed out that
\begin{quote}
Possible implications of dynamic allocation methods [minimization] on the analysis e.g. with regard to bias and Type I error control should be carefully considered,... it has been shown that in case of dynamic treatment allocation conventional statistical methods do not always control the Type I error.
\end{quote}

To overcome the shortcomings, we propose a new minimization procedure, which can be viewed as an improvement of PS procedure. The proposed procedure has two main advantages. First, the marginal imbalances are bounded in probability, as under the PS procedure. Second, under the new procedure, the imbalance of additional covariates, including within-stratum imbalance, is asymptotically normal with asymptotic variance no larger than that under CR, which implies that neither the shift problem nor variance inflation arises under the new procedure.

The remainder of this article is organized as follows. In Section \ref{Sec2}, we describe the proposed procedure. In Section \ref{Sec3}, we establish the theoretical properties. In Sections \ref{Sec4} and \ref{Sec5}, we present simulation studies and a clinical trial example, respectively. In Section \ref{Sec6}, we summarize the study and provide directions for future work. Technical proofs are provided in the appendix.

\section{New Minimization Procedure}\label{Sec2}
Consider a two-arm trial in which units are divided into a treatment group and a control group, with the target allocation proportion assigned to the treatment group denoted by $\pi\in(0,1)$. Let $T_i$ denote the assignment of the $i$th unit, where $T_i=1$ indicates assignment to the treatment group and $T_i=0$ indicates assignment to the control group. Let $X_i = (X_{i,1},\cdots,X_{i,p})^\top$ denote a $p$-dimensional vector of discrete covariates, with $X_{i,j}\in\{1,\cdots,m_j\}$. If a covariate is continuous, its discretized version is used instead. Then the new procedure is defined as follows.\\
Step 1: The first unit is assigned to the treatment group with probability $\pi$ and to the control group with probability $1-\pi$. \\
Step 2: Suppose that $k-1$ units have been assigned. We calculate every marginal imbalance $D_{k-1}(j;s_j) = \sum_{i=1}^{k-1} (T_i-\pi)\mathbb{I}\{X_{i,j} = s_j\} $ and choose
\begin{align*}
(j^*;s_j^*) \in \arg\max_{j, s_j } \left| D_{k-1}(j;s_j) \right|.
\end{align*}
Step 3: The $k$th unit is assigned to the treatment group according to the following allocation rule.
\begin{align*}
 \mathbb{P}\left(T_k = 1 \mid \mathcal{G}_{k-1},X_k,j^*,s_j^* \right) 
= 
\begin{cases}
	\rho & \text{if }  X_{k,j^*} = s_j^* \text{ and } D_{k-1}(j^*;s_j^*) < 0 ,  \\
	2\pi - \rho & \text{if } X_{k,j^*} = s_j^* \text{ and } D_{k-1}(j^*;s_j^*) > 0, \\
	\pi & \text{otherwise} ,
\end{cases}
\end{align*}
where $\pi < \rho < \min\{1,2\pi \} $ and $\mathcal{G}_{k-1}$ denotes the $\sigma$-field generated by $\{T_i,X_i\}_{i=1}^{k-1}$.  \\
Step 4: Repeat Steps 2-3 until all units are assigned. 

\begin{remark}
In Step 2, when multiple coordinates attain this maximum, a coordinate is typically selected uniformly at random. Also, the choice can be determined using prespecified weights that reflect the experimenter's priorities, or deterministically according to a prespecified ordering. However, the selection rule should be fixed in advance and must not depend on the next unit's covariates.
\end{remark}

\section{Theoretical Properties}\label{Sec3}

In this section, we investigate the convergence rate of the marginal imbalance as well as the asymptotic properties of $n^{-1/2}\sum_{i=1}^{n} (T_i-\pi)Z_i $ for arbitrary observed or unobserved covariates $\{Z_i\}_{i=1}^n$. We begin with the following assumption.  
\begin{assumption}\label{assump_iid}
The covariates $\{(X_i^\top,Z_i)^\top \}_{i=1}^n$ are independent and identically distributed copies of $(X^\top,Z)^\top$ with $X=(X_1,\cdots,X_p)^\top$ and $\mathbb{E}[Z^2] < \infty$. 
\end{assumption}

\begin{theorem}\label{Thm_Lambda}
Suppose that Assumption \ref{assump_iid} holds. Then, $D_n(j;s_j) = O_P(1)$ for $j=1,\cdots,p$ and $s_j = 1,\cdots,m_j$. 
\end{theorem}

Theorem \ref{Thm_Lambda} implies that, under the proposed CAR procedure, the marginal imbalances are bounded in probability under both equal and unequal allocation, matching that of the marginal imbalance under the PS procedure (\citealp{hu2020theory}). In addition, as a direct corollary, the overall imbalance $D_n = \sum_{i=1}^{n} (T_i-\pi)$ is also bounded in probability, as it equals the sum of the marginal imbalances across the levels of any one covariate.

Next, we establish the asymptotic behavior of $n^{-1/2}\sum_{i=1}^{n} (T_i-\pi)Z_i $. 
\begin{theorem}\label{Thm_CLT}
Suppose that Assumption \ref{assump_iid} holds. Let $\mathcal{X} = \mathrm{span}\{\mathbb{I}\{X_{j} = s_j\}: j=1,\cdots,p; s_j = 1,\cdots,m_j  \}$ denote the linear space spanned by the level-indicator variables of all covariates. Then,
\begin{align*}
	\frac{1}{\sqrt{n}} \sum_{i=1}^{n} (T_i-\pi)Z_i \xrightarrow{d} \mathcal{N}\left(0,\pi(1-\pi)\left\{\mathbb{E}\left[Z^2\right] - \mathbb{E}\left[(\Pi_{\mathcal{X}} Z)^2 \right]\right\} \right),
\end{align*}
where $\Pi_{\mathcal{X}} Z$ denotes the linear projection of $Z$ onto $\mathcal{X}$, that is, $\Pi_{\mathcal{X}} Z = \arg\min_{W\in\mathcal{X}} \mathbb{E}[(Z-W)^2]$. 
\end{theorem}
Theorem \ref{Thm_CLT} shows two significant advantages of our proposed procedure. First, under unequal allocation, the proposed procedure eliminates the shift problem, which refers to the phenomenon that for some additional covariate $Z_i$ not used in the procedure, $n^{-1} \sum_{i=1}^n (T_i-\pi)Z_i \xrightarrow{P} c_Z \ne 0$ as $n \to \infty$ when $\pi\ne 1/2$, and this phenomenon can occur under PS and other CAR procedures (\citealp{liu2025properties}, \citealp{fang2026general}). Second, the asymptotic variance is ensured to be no larger than that under CR, a guarantee that does not hold in general for the PS procedure (\citealp{zhao2024consistent}).

\begin{remark}
For the proposed procedure, as with the PS procedure, the vector $\Lambda_n = (D_{n}(j;s_j))_{j=1,\cdots,p;\, s_j=1,\cdots,m_j}$, which consists of all marginal imbalances, is a time-homogeneous Markov chain. However, to accommodate irrational values of $\pi$, we do not follow the conventional approach of \cite{HuHu} and \cite{hu2020theory} in our proofs, which establishes boundedness in probability through positive recurrence of $\{\Lambda_n\}$.
\end{remark}

\begin{remark}
Following the discovery of the shift problem under unequal allocation by \citet{liu2025properties}, \cite{zhang2026covariate} and \cite{fang2026general} proposed two different CAR procedures to solve the problem within the unified framework, respectively. Both procedures depart from the classical biased-coin allocation probability in CAR. For the procedure of \cite{zhang2026covariate}, which reduces to the selection bias reduced CAR procedure (\citealp{zhang2026efficient}) when solely considering the marginal imbalance, the asymptotic variance of the additional covariate imbalances is the same as that in Theorem \ref{Thm_CLT}, but the convergence rate of marginal imbalance is slower than the proposed procedure. For the procedure of \citet{fang2026general}, it has not been established that the asymptotic variance of additional covariate imbalances is no larger than that under CR. The proposed procedure keeps the classical biased-coin allocation probability, while achieving bounded marginal imbalances with guarantees against both the shift problem and variance inflation.
\end{remark}

\section{Simulation}\label{Sec4}
In this section, we evaluate the proposed procedure through simulations to illustrate the theoretical results. We compare the new procedure with complete randomization and Pocock and Simon's minimization procedure. Under unequal allocation, we consider the PS procedure extended by \cite{liu2017general}, which is also a special case of the framework in \cite{liu2025properties}. Throughout the simulation studies, when multiple candidates attain the largest absolute marginal imbalance in Step 2 of the proposed procedure, the choice is made uniformly at random. Unless otherwise specified, the biased coin probability is set to $\rho=0.9$ for both PS and the proposed procedure.

\subsection{Convergence Rate of Marginal Imbalance}\label{sec_4.1}
First, we evaluate the convergence rate of marginal imbalance of the new procedure and compare the results with those under CR and PS. The covariates $X_i$ are two-dimensional, where $X_{i,1}$ takes values in $\{1,2\}$ with equal probability and $X_{i,2}$ takes values in $\{1,2\}$ with probabilities of 0.7 and 0.3, respectively. The two covariates are independent of each other. The sample sizes are $n\in\{100,200,400,800\}$ and each case is based on 5000 replicates. Table \ref{table1} reports the simulated standard deviations of $D_n(1;1)$ and $D_n(2;1)$ for CR, PS and the new procedure under equal allocation ($\pi=1/2$) and unequal allocation ($\pi=2/3$).

Table \ref{table1} suggests that all standard deviations of the marginal imbalances stabilize under the PS procedure and the new procedure, which is consistent with the $O_P(1)$ convergence rate established theoretically. Under CR, however, the standard deviations increase as the sample size grows. Moreover, although the PS procedure and the new procedure share the same convergence rate, the standard deviations of the marginal imbalances under the PS procedure are all smaller than those under the new procedure, indicating that the PS procedure yields less covariate imbalance. 

\begin{table}[H]
	\centering
	\caption{Standard deviations of selected marginal imbalances of several randomization procedures. }
	\label{table1}
	\begin{tabular}{*{6}{c}}
		\toprule
		& & \multicolumn{2}{c}{$\pi=1/2$} & \multicolumn{2}{c}{$\pi=2/3$} \\
		\cmidrule(lr{0pt}){3-4} \cmidrule(lr{0pt}){5-6}
		Randomization & $n$ & $D_n(1;1)$ & $D_n(2;1)$ & $D_n(1;1)$ & $D_n(2;1)$ \\
		\midrule
		CR & 100 & 3.55 & 4.17 &  3.32& 4.04   \\
		   & 200 & 4.91 & 5.86 &  4.71&  5.52  \\
		   & 400 & 7.13 & 8.44 &  6.61&  7.90  \\
		   & 800 & 9.86 & 11.74 &  9.54& 11.21 \\
		PS & 100 & 0.50 & 0.49&  0.53& 0.52    \\
		   & 200 & 0.49 & 0.49& 0.52& 0.52     \\
		   & 400 &  0.51& 0.50&  0.53& 0.52    \\
		   & 800 &  0.50& 0.50&  0.51& 0.52    \\
	   New & 100 & 0.75 & 0.76 &  1.14& 1.15   \\
		   & 200 & 0.79 & 0.77 &  1.15& 1.17   \\
		   & 400 & 0.77 & 0.76 &  1.16& 1.16   \\
		   & 800 & 0.76 & 0.75 &  1.15& 1.13   \\
		\bottomrule
	\end{tabular}
\end{table}

\subsection{No Shift Problem}
\begin{figure}[H]
	\centering
	\includegraphics[width=\textwidth]{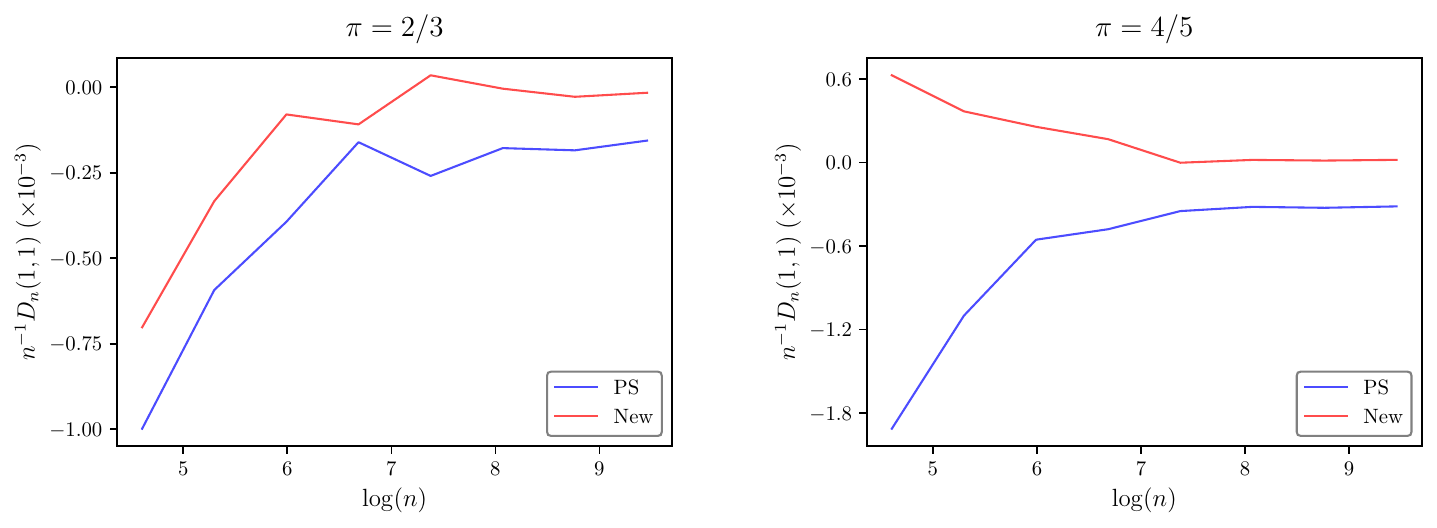}
	\caption{Means of $n^{-1}D_n(1,1) $ with different sample sizes under PS and the new procedure} 
	\label{fig1}
\end{figure}

Second, we want to show that our proposed procedure has no ``shift problem" under unequal allocation. The covariates $X_i$ are two-dimensional, where $X_{i,1}$ takes values in $\{1,2\}$ with probabilities of 0.8 and 0.2 and $X_{i,2}$ takes values in $\{1,2\}$ with probabilities of 0.7 and 0.3, respectively. The two covariates are independent of each other. The sample sizes are $n\in\{100,200,400,800,1600,3200,6400,12800\}$ and each case is based on 5000 replicates. Figure \ref{fig1} presents the simulation results for the PS procedure and the new procedure when $\pi=2/3$ and $\pi =4/5$, where the horizontal axis is $\log(n)$ and the vertical axis is the average of $n^{-1}D_n(1,1) = n^{-1} \sum_{i=1}^{n} (T_i-\pi)\mathbb{I}\{X_{i,1}=X_{i,2}=1\}$ over 5000 replications.
In Figure \ref{fig1}, means of $n^{-1}D_n(1,1)$ converge to a non-zero limit under PS as the sample size increases, whereas they converge to zero under the proposed procedure. This contrast indicates that PS suffers from the shift problem, while the proposed procedure avoids the shift problem observed under PS. 

\subsection{No Variance Inflation}
\begin{figure}[H]
	\centering
	\includegraphics[width=\textwidth]{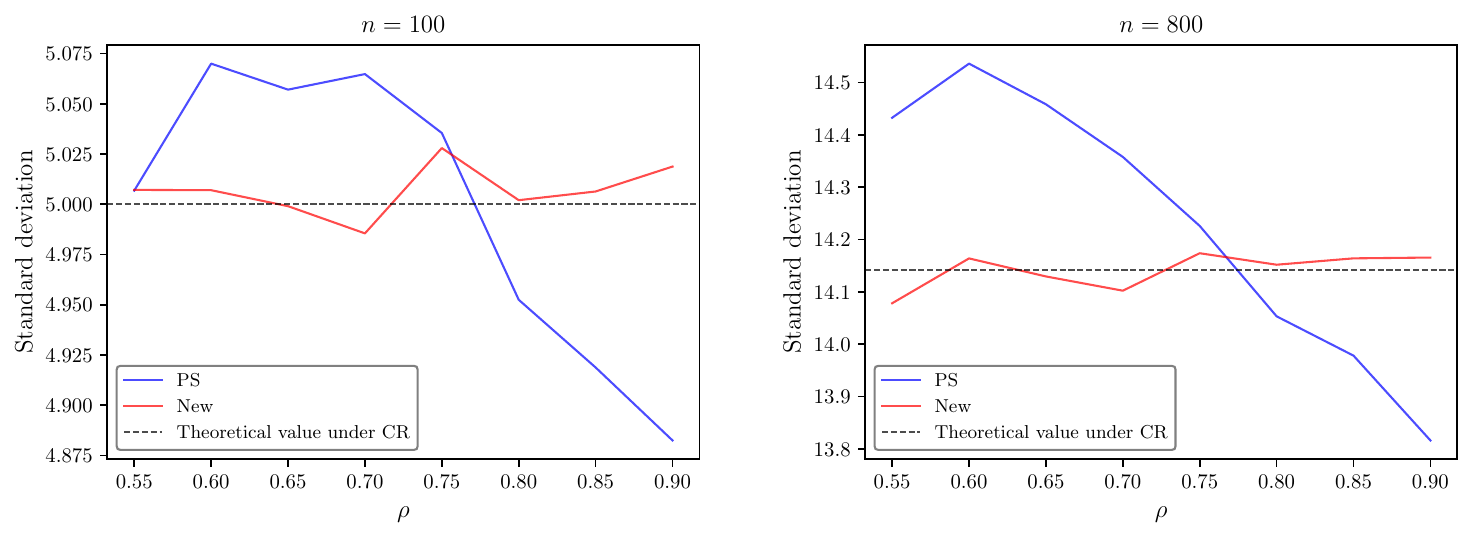}
	\caption{Standard deviations of $D_n(1,1)-D_n(1,2)-D_n(2,1)+D_n(2,2)$ with different biased coin probabilities under PS and the new procedure}
	\label{fig2}
\end{figure}

Third, we want to show that our proposed procedure has no inflated variance and the evaluation is under equal allocation and based on the counterexample provided in \cite{zhao2024consistent}, which aimed to show that PS procedure may cause inflated variance. To be specific, the covariates $X_i$ are two-dimensional with two covariates being independent, all taking values in $\{1,2\}$ with equal probability. The biased coin probability $\rho\in\{0.55,0.6,0.65,\cdots,0.9\}$. The sample sizes are $n\in\{100,800\}$ and each case is based on 50000 replicates. Figure \ref{fig2} presents the simulation results for the PS procedure and the new procedure, where the horizontal axis is $\rho$ and the vertical axis is the standard deviations of $D_n(1,1)-D_n(1,2)-D_n(2,1)+D_n(2,2)$ over 50000 replications. Here, $D_n(a,b) = \sum_{i=1}^n (T_i-1/2)\mathbb{I}\{X_{i,1}=a,X_{i,2}=b\}$ for $a,b\in\{1,2\}$ denotes the within-stratum imbalance in stratum $(a,b)$. In addition, we add a black dashed line to the figure, which represents the theoretical value under complete randomization, which is also the theoretical value under our proposed procedure. 
In Figure \ref{fig2}, when $\rho\leq 0.75$, the standard deviations under PS are higher than the theoretical value under CR, while when $\rho \geq 0.8$, the standard deviations are lower than the theoretical value under CR. In contrast, the standard deviations under the proposed procedure remain nearly constant as $\rho$ varies and stay close to their theoretical values. This phenomenon is particularly pronounced at a sample size of 800. These results suggest that the PS procedure does not guarantee the absence of variance inflation, while the proposed procedure provides such a guarantee.

\section{Real Data Example}\label{Sec5}
In this section, we compare the performance of the proposed procedure with CR and PS through the re-design of ACTG-320 study (\citealp{hammer1997controlled}), which was a randomized, double-blind, placebo-controlled clinical trial involving 1,156 HIV-infected patients. It aimed to compare the outcomes of a three-drug regimen (indinavir, zidovudine, and lamivudine) with those of a two-drug regimen (zidovudine and lamivudine). We consider four covariates and the detailed information is shown in Table \ref{table_realdata_covariates}.

\begin{table}[H]
	\centering
	\small
	\caption{Description of the selected covariates in the ACTG-320 study. }
	\label{table_realdata_covariates}
	\begin{tabular}{*{3}{l}}
	\toprule
	Covariate & Description & Levels and stratification \\
	\midrule
	Sex & Sex of participant & 2 levels: Male, Female \\
	CD4 & CD4 count at screening & 2 levels: $\leq 50$, $>50$  \\
	Karnof & Karnofsky performance score & 4 levels: 70, 80, 90, 100 \\ 
	ZDV & Duration of prior zidovudine use (months) & 4 levels: $\leq 10$, $(10,20]$, $(20,40]$, $>40$  \\
	\bottomrule	
    \end{tabular}
\end{table}
We assess the balancing performance of CR, PS, and the proposed procedure for the four selected covariates. For each target allocation proportion, $\pi\in\{1/2,2/3\}$, we repeat the randomization 5,000 times. Table \ref{table_realdata} reports the standard deviations of the selected marginal imbalance $D_n$(Sex; Male), $D_n$(CD4; $\leq 50$), $D_n$(Karnof; 70), $D_n$(ZDV; $\leq 10$) together with that of the within-stratum imbalance $D_n(s)$ for the stratum $s=(\text{Sex}=\text{Male},\text{CD4}\leq 50,\text{Karnof}=70,\text{ZDV}\leq 10)$.

Under both allocation ratios, PS and the new procedure yield smaller standard deviations for all four selected marginal imbalances than CR. In addition, PS consistently yields smaller marginal imbalance than the proposed procedure, while the three procedures exhibit similar within-stratum imbalances. This is consistent with the findings from our simulation studies.

\begin{table}[H]
	\centering
	\small
	\caption{The ACTG-320 study: standard deviations of selected marginal imbalances and within-stratum imbalance of several randomization procedures. }
	\label{table_realdata}
	\begin{tabular}{*{7}{c}}
		\toprule
	$\pi$& 	Ran. & $D_n$(Sex; Male) & $D_n$(CD4; $\leq 50$) & $D_n$(Karnof; 70) & $D_n$(ZDV; $\leq 10$) & $D_n(s)$ \\
		\midrule
  $1/2$  &	CR & 15.20 & 10.34 & 2.82 & 8.61 & 0.71  \\
		 &  PS & 0.68 & 0.66  & 0.70 & 0.70 & 0.70 \\
		 & New & 1.74 & 1.74  & 1.76  & 1.78 & 0.70 \\
		\midrule
  $2/3$	&	CR & 14.58 & 9.88& 2.63& 8.18& 0.67  \\
		&  PS & 0.68 & 0.69& 0.71& 0.73& 0.65 \\
		& New & 2.48 & 2.49&  2.17& 2.59& 0.68 \\
		\bottomrule
	\end{tabular}\\[4pt]
\begin{minipage}{\linewidth}
	\raggedright\noindent Abbreviation: Ran., Randomization.  
\end{minipage}
\end{table}

\section{Conclusion}\label{Sec6}
In this paper, we propose a new minimization procedure, which can be viewed as an improvement of Pocock and Simon's minimization procedure. This procedure keeps marginal imbalances bounded in probability and avoids the shift problem and variance inflation, which are established theoretically and supported by simulation studies. The results of this article can be further extended in several directions. First, the proposed framework could be generalized to accommodate multiple treatments (\citealp{hu2023multi}). Second, the theoretical properties of the procedure under high-dimensional covariates, and its comparison with the PS procedure, are also important directions for future research (\citealp{tao2026theoretical}). Finally, it would be valuable to extend the proposed procedure into a unified framework that accommodates a broader class of feature maps and imbalance measures (\citealp{ma2024new}, \citealp{zhang2023asymptotic}, \citealp{liu2025properties}).

\section*{Acknowledgments}
This work was supported by  grants from National Key R\&D Program of China (No. 2024YFA1013502), NSF of China (Grant No. U23A2064) and the Summit Advancement Disciplines of Zhejiang Province (Zhejiang Gongshang University - Statistics).

\nocite{*}
\bibliographystyle{chicago} 
\bibliography{ref.bib}

\appendix
\newpage
\bigskip
\begin{center}
{\large\bf SUPPLEMENTARY MATERIAL}
\end{center}

\setcounter{equation}{0}
\numberwithin{equation}{section}
\numberwithin{table}{section}
\numberwithin{figure}{section}



\section{Proof of the Main Results}
\begin{proof}[Proof of Theorem \ref{Thm_Lambda}:] Write $\phi(X_i) = (\mathbb{I}\{X_{i,j}= s_j\})_{1\leq j \leq p;\, 1\leq s_j \leq m_j}^\top\in \{0,1\}^m$, where $m =\sum_{j=1}^p m_j$ and $\Lambda_k = \sum_{i=1}^{k} (T_i-\pi)\phi(X_i) = (D_{k}(j;s_j))_{j=1,\cdots,p;\, s_j=1,\cdots,m_j}$. Let $J:\mathbb{R}^m \times  (0,1) \mapsto \{1,\cdots,m\}$ be a fixed measurable selection function satisfying
\begin{align*}
J(\Lambda,U) \in \arg\max_{1\leq j \leq m} \left| e_j^\top\Lambda\right|,
\end{align*}
where $e_j$ denotes the $j$-th standard basis vector of $\mathbb{R}^m$. Set $J_k=J(\Lambda_k,U_k)$, where $U_k\sim\operatorname{Unif}(0,1)$
is an auxiliary random variable independent of all covariates information. The function $J$ implements the pre-specified selection rule, so that $J_k$ is determined by $\Lambda_k$ and $U_k$. Let $\mathcal{F}_k$ denote the $\sigma$-field generated by $\{T_i,X_i,U_i,Z_i\}_{i=1}^k$. Then, $\Lambda_k$ and $J_k$ are $\mathcal{F}_k$-measurable and 
\begin{align*}
\mathbb{E}\left[(T_{k+1} - \pi) \,\middle|\, \mathcal{F}_k, X_{k+1} \right] = -(\rho-\pi)\, \mathrm{sgn}\left\{ \Lambda_{k,J_k} \right\} \phi(X_{k+1})^\top e_{J_k}. 
\end{align*}
It follows that
\begin{align*}
\mathbb{E}\left[(T_{k+1} - \pi)\phi(X_{k+1}) \,\middle|\, \mathcal{F}_k, X_{k+1} \right] = -(\rho-\pi)\, \mathrm{sgn}\left\{ \Lambda_{k,J_k} \right\} \phi(X_{k+1}) \phi(X_{k+1})^\top e_{J_k},
\end{align*}	
which implies that 
\begin{align*}
\mathbb{E}\left[(T_{k+1} - \pi)\phi(X_{k+1}) \,\middle|\, \mathcal{F}_k \right] =  -(\rho-\pi)\, \mathrm{sgn}\left\{ \Lambda_{k,J_k} \right\} \boldsymbol{\Sigma}_{\phi\phi}  e_{J_k},
\end{align*}
where $\boldsymbol{\Sigma}_{\phi\phi} = \mathbb{E}[\phi(X)\phi(X)^\top]$. \\
Define $V_k = V(\Lambda_k) =\Lambda_k^\top \boldsymbol{\Sigma}_{\phi\phi}^+ \Lambda_k$, where $\boldsymbol{\Sigma}_{\phi\phi}^+$ denotes the Moore–Penrose inverse of $\boldsymbol{\Sigma}_{\phi\phi}$. Then,
\begin{align*}
&\ \mathbb{E}\left[V_{k+1} \,\middle|\, \mathcal{F}_k   \right] - V_{k} \\
=&\ 2 \Lambda_{k}^\top \boldsymbol{\Sigma}_{\phi\phi}^+ \mathbb{E}\left[(T_{k+1} - \pi)\phi(X_{k+1}) \,\middle|\, \mathcal{F}_k \right]  + \mathbb{E}\left[(T_{k+1}-\pi)^2 \phi(X_{k+1})^\top \boldsymbol{\Sigma}_{\phi\phi}^+ \phi(X_{k+1}) \,\middle|\,  \mathcal{F}_k \right] \\
\leq&\ -2 (\rho-\pi)\, \mathrm{sgn}\left\{ \Lambda_{k,J_k} \right\} \Lambda_{k}^\top \boldsymbol{\Sigma}_{\phi\phi}^+ \boldsymbol{\Sigma}_{\phi\phi}  e_{J_k} + \mathbb{E}\left[ \phi(X)^\top \boldsymbol{\Sigma}_{\phi\phi}^+ \phi(X) \right] \\
=&\ -2(\rho-\pi) \left\|\Lambda_{k} \right\|_{\infty} + \mathrm{rank}(\boldsymbol{\Sigma}_{\phi\phi}),
\end{align*}
where the last equation follows from the fact that
\begin{align*}
\mathrm{sgn}\left\{ \Lambda_{k,J_k} \right\} \Lambda_{k}^\top   e_{J_k} = \mathrm{sgn}\left\{ \Lambda_{k,J_k} \right\} \Lambda_{k,J_k} = \left|  \Lambda_{k,J_k} \right| = \left\|\Lambda_{k} \right\|_{\infty}. 
\end{align*}
By an elementary inequality that for any $x>0$ and $y\geq 0$, 
\begin{align*}
	\sqrt{y} - \sqrt{x} \leq \frac{y-x}{2\sqrt{x}}. 
\end{align*}
Then, if $V_k >0$, 
\begin{align*}
	\mathbb{E}\left[V_{k+1}^{1/2} \,\middle|\, \mathcal{F}_k  \right] - V_{k}^{1/2} \leq -(\rho-\pi)\, \frac{\left\|\Lambda_{k} \right\|_{\infty}}{V_{k}^{1/2}} + \frac{\mathrm{rank}(\boldsymbol{\Sigma}_{\phi\phi})}{2V_{k}^{1/2}}.  
\end{align*}
Note that $\left\|\Lambda_{k} \right\|_{\infty} \geq m^{-1/2} \left\|\Lambda_k\right\|_2$ and $V_{k}^{1/2} \leq \lambda_{0}^{-1/2} \left\| \Lambda_k \right\|_2$, where $\lambda_{0}$ denotes the minimum nonzero eigenvalue of $\boldsymbol{\Sigma}_{\phi\phi}$. Hence, 
\begin{align*}
	\mathbb{E}\left[V_{k+1}^{1/2} \,\middle|\, \mathcal{F}_k  \right] - V_{k}^{1/2} \leq -(\rho-\pi) \left(\frac{\lambda_{0}}{m}\right)^{1/2} + \frac{\mathrm{rank}(\boldsymbol{\Sigma}_{\phi\phi})}{2V_{k}^{1/2}},
\end{align*}
which implies that there exist positive constants $c$ and $R$, such that on $\mathcal{C}_k = \{ V_{k}^{1/2} \geq R \}$, 
\begin{align}\label{proof_eq1}
	\mathbb{E}\left[V_{k+1}^{1/2} \,\middle|\, \mathcal{F}_k  \right] - V_{k}^{1/2} \leq -c . 
\end{align}
In addition,
\begin{align*}
	&\ \left| V_{k+1}^{1/2}  - V_{k}^{1/2} \right| = \left|\, \left\| \left(\boldsymbol{\Sigma}_{\phi\phi}^+ \right)^{1/2} \Lambda_{k+1} \right\|_2 - \left\| \left(\boldsymbol{\Sigma}_{\phi\phi}^+ \right)^{1/2} \Lambda_{k} \right\|_2 \,\right| \\
	\leq&\ \left\| \left(\boldsymbol{\Sigma}_{\phi\phi}^+ \right)^{1/2}  (T_{k+1}-\pi) \phi(X_{k+1}) \right\|_2 \leq \left\| \left(\boldsymbol{\Sigma}_{\phi\phi}^+ \right)^{1/2} \phi(X_{k+1}) \right\|_2 = \left\{ \phi(X_{k+1})^\top \boldsymbol{\Sigma}_{\phi\phi}^+ \phi(X_{k+1}) \right\}^{1/2} \\
	\leq&\ \lambda_0^{-1/2} \left\|\phi(X_{k+1})\right\|_2 = \lambda_0^{-1/2}p^{1/2}, 
\end{align*}
which implies that for any $r\geq 1$,
\begin{align}\label{proof_eq2}
\mathbb{E}\left[ \left|V_{k+1}^{1/2} - V_k^{1/2} \right|^r \,\middle|\, \mathcal{F}_k \right] \leq \lambda_0^{-r/2}p^{r/2}. 
\end{align}
By (\ref{proof_eq1}), (\ref{proof_eq2}) and Theorem 1 of \cite{pemantle1999moment}, we have $\mathbb{E}[ V_n^{1/2}] \leq C_V$ for some constant $C_V>0$, which implies that
\begin{align*}
\mathbb{E}\left[ \| \Lambda_n\|_2 \right] \leq  \lambda_{\mathrm{max}}^{1/2} \mathbb{E}\left[ V_n^{1/2}\right] \leq  \lambda_{\mathrm{max}}^{1/2} C_V,
\end{align*}
where $\lambda_{\mathrm{max}}$ denotes the maximal eigenvalues of $\boldsymbol{\Sigma}_{\phi\phi}$. Therefore, by Markov's inequality, $\Lambda_n = O_P(1)$.
\end{proof}

\begin{proof}[Proof of Theorem \ref{Thm_CLT}: ]
Consider the following decomposition,
\begin{align*}
(T_i-\pi)Z_i = (T_i-\pi)\left\{Z_i-\beta_Z^\top\phi(X_i)\right\} +  (T_i-\pi)\beta_Z^\top\phi(X_i), 
\end{align*}
where $\beta_Z = \boldsymbol{\Sigma}_{\phi\phi}^+ \mathbb{E}[\phi(X)Z]$. \\
Note that $\mathbb{E}[\phi(X)\{ Z-\beta_Z^\top\phi(X) \}] = \mathbb{E}[\phi(X)Z] - \boldsymbol{\Sigma}_{\phi\phi} \beta_Z =0 $. Then,  
\begin{align*}
&\ \mathbb{E}\left[(T_i-\pi)\left\{Z_i-\beta_Z^\top\phi(X_i)\right\} \,\middle|\, \mathcal{F}_{i-1} \right] \\
=&\ -(\rho-\pi)\, \mathrm{sgn}\left\{ \Lambda_{i-1,J_{i-1}} \right\} \mathbb{E}\left[\left\{Z_i-\beta_Z^\top\phi(X_i)\right\} \phi(X_{i})^\top\right] e_{J_{i-1}} = 0.
\end{align*}
It follows that $\{ (T_i-\pi)\{Z_i-\beta_Z^\top\phi(X_i)\},\mathcal{F}_i  \}_{i=1}^n$ is a sequence of martingale difference and 
\begin{align*}
\sum_{i=1}^n (T_i-\pi)Z_i &= \sum_{i=1}^n (T_i-\pi)\left\{Z_i-\beta_Z^\top\phi(X_i)\right\} + \beta_Z^\top \Lambda_n \\
&= \sum_{i=1}^n (T_i-\pi)\left\{Z_i-\beta_Z^\top\phi(X_i)\right\} + O_P(1). 
\end{align*}
A direct calculation yields that
\begin{align*}
&\ \frac{1}{n} \sum_{i=1}^n \mathbb{E}\left[ (T_i-\pi)^2 \left\{Z_i-\beta_Z^\top\phi(X_i)\right\}^2 \,\middle|\, \mathcal{F}_{i-1} \right] \\
=&\ \frac{1}{n} \sum_{i=1}^n (1-2\pi)\mathbb{E}\left[ (T_i-\pi) \left\{Z_i-\beta_Z^\top\phi(X_i)\right\}^2 \,\middle|\, \mathcal{F}_{i-1} \right] + \pi(1-\pi) \mathbb{E}\left[ \left\{Z-\beta_Z^\top\phi(X)\right\}^2  \right] \\
=&\ -(\rho-\pi)(1-2\pi) \mathbb{E}\left[\left\{Z-\beta_Z^\top\phi(X)\right\}^2 \phi(X)^\top\right] \frac{1}{n} \sum_{i=1}^n \mathrm{sgn}\left\{ \Lambda_{i-1,J_{i-1}} \right\}  e_{J_{i-1}}\\
&\ + \pi(1-\pi) \mathbb{E}\left[ \left\{Z-\beta_Z^\top\phi(X)\right\}^2  \right]. 
\end{align*}
Note that
\begin{align*}
\Lambda_n &= \sum_{i=1}^n \left\{(T_i-\pi)\phi(X_i) - \mathbb{E}\left[ (T_i-\pi)\phi(X_i)\,\middle|\, \mathcal{F}_{i-1}  \right] \right\} + \sum_{i=1}^n \mathbb{E}\left[ (T_i-\pi)\phi(X_i)\,\middle|\, \mathcal{F}_{i-1}  \right] \\
&= -(\rho-\pi) \boldsymbol{\Sigma}_{\phi\phi} \sum_{i=1}^n  \mathrm{sgn}\left\{ \Lambda_{i-1,J_{i-1}} \right\}   e_{J_{i-1}} + o_P(n),
\end{align*}
where the last equation follows from the law of large numbers for martingale difference sequence.\\
It follows that
\begin{align*}
\frac{1}{n}\boldsymbol{\Sigma}_{\phi\phi} \sum_{i=1}^n  \mathrm{sgn}\left\{ \Lambda_{i-1,J_{i-1}} \right\}   e_{J_{i-1}} \xrightarrow{P} 0. 
\end{align*}
In addition, for any vector $u\in \operatorname{ker}(\boldsymbol{\Sigma}_{\phi\phi}) = \{v: \boldsymbol{\Sigma}_{\phi\phi} v =0  \}$, we have $\mathbb{E}[|u^\top \phi(X)|^2] = u^\top \boldsymbol{\Sigma}_{\phi\phi} u = 0$, which implies that $u^\top \phi(X) = 0$ almost surely. Hence, 
\begin{align*}
u^\top \mathbb{E}\left[\left\{Z-\beta_Z^\top\phi(X)\right\}^2 \phi(X) \right] = 0,
\end{align*}
which implies that $\mathbb{E}[\{Z-\beta_Z^\top\phi(X)\}^2 \phi(X)] \in \mathrm{range}( \boldsymbol{\Sigma}_{\phi\phi} ) = \{ \boldsymbol{\Sigma}_{\phi\phi}v:v\in\mathbb{R}^m \}$.\\
Therefore, we obtain that
\begin{align*}
\frac{1}{n}\, \mathbb{E}\left[\left\{Z-\beta_Z^\top\phi(X)\right\}^2 \phi(X)^\top \right] \sum_{i=1}^n  \mathrm{sgn}\left\{ \Lambda_{i-1,J_{i-1}} \right\}   e_{J_{i-1}} \xrightarrow{P} 0,
\end{align*}
which implies that
\begin{align*}
\frac{1}{n} \sum_{i=1}^n \mathbb{E}\left[ (T_i-\pi)^2 \left\{Z_i-\beta_Z^\top\phi(X_i)\right\}^2 \,\middle|\, \mathcal{F}_{i-1} \right] \xrightarrow{P} \pi(1-\pi) \mathbb{E}\left[ \left\{Z-\beta_Z^\top\phi(X)\right\}^2  \right]. 
\end{align*}
Furthermore, Lindeberg's condition is satisfied. In fact, for any $\varepsilon>0$,
\begin{align*}
&\ \frac{1}{n} \sum_{i=1}^{n} \mathbb{E}\left[  (T_i-\pi)^2 \left\{Z_i-\beta_Z^\top\phi(X_i)\right\}^2 \mathbb{I}\left\{ \left| (T_i-\pi) \left\{Z_i-\beta_Z^\top\phi(X_i)\right\} \right| > \varepsilon \sqrt{n}\right\}\,  \middle|\, \mathcal{F}_{i-1} \right] \\
\leq&\ \frac{1}{n} \sum_{i=1}^n  \mathbb{E}\left[ \left\{Z_i-\beta_Z^\top\phi(X_i)\right\}^2 \mathbb{I}\left\{ \left|Z_i-\beta_Z^\top\phi(X_i) \right| > \varepsilon \sqrt{n}\right\}\,  \middle|\, \mathcal{F}_{i-1} \right] \\
=&\ \mathbb{E}\left[ \left\{Z-\beta_Z^\top\phi(X)\right\}^2 \mathbb{I}\left\{ \left|Z-\beta_Z^\top\phi(X) \right| > \varepsilon \sqrt{n}\right\} \right] \xrightarrow{P} 0.
\end{align*}
Therefore, by the martingale central limit theorem and noting that $\Pi_\mathcal{X} Z = \beta_Z^\top \phi(X)$, we finish the proof of the theorem. 
\end{proof}

\end{document}